# The Evolution and Current State of Swift-Heavy-Ion Track Modeling

N. Medvedev[1,2,*], A.E. Volkov[3]

*1) Institute of Physics, Czech Academy of Sciences, Na Slovance 1999/2, 182 00 Prague 8, Czechia*

*2) Institute of Plasma Physics, Czech Academy of Sciences, U Slovanky 2525/1a, 182 00 Prague 8, Czechia*

*3) P.N. Lebedev Physical Institute of the Russian Academy of Sciences, Leninskij pr., 53,119991 Moscow, Russia*

## Abstract

This paper briefly reviews the modeling approaches to the swift heavy ion (SHI) track formation. It presents a step-by-step derivation of the two-temperature model, highlighting its underlying assumptions and approximations. Special attention is given to identifying assumptions that fail under the typical SHI track conditions, and what methods currently exist to go beyond them. As an alternative to analytical approximations, a combined Monte Carlo-molecular dynamics model is demonstrated, outlining its basic assumptions, their justifications, and validations. Finally, the paper addresses bottlenecks of the current multiscale approach, its remaining weaknesses, and necessary further developments of the theory and models.

## Introduction

Understanding the basic mechanisms of interaction of swift ions with matter is of paramount practical importance: from understanding the dangers to biological cells exposed to cosmic rays or hadron treatment, to single event upsets in electronics, to design of nano-technology, such as nano-membranes or quantum dots [1–6]. An impinging ion induced a multitude of intertwined effects in the target matter, ranging over many orders of magnitude in time and space [7–9], see illustration in Figure 1. The track of modified material left behind an ion typically ranges from microns to millimetres in length, while affecting only a few nanometres in diameter. Point defects and dislocations may appear in a larger radius up to a micron [7,8].

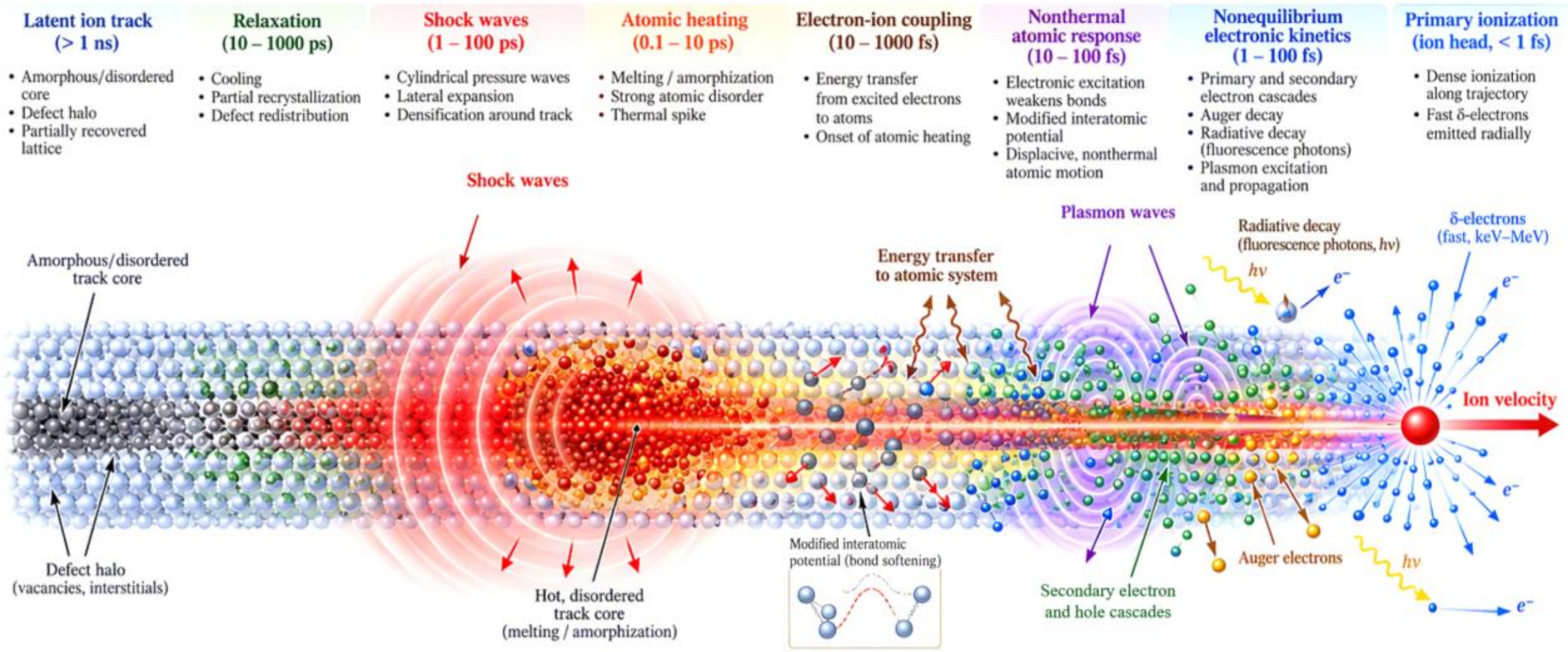

*Figure 1. Schematic illustration of the processes induced by a swift heavy ion in matter, created by Microsoft Copilot AI, prompt based on the review [9].*

A swift heavy ion (SHI) with energies above ~MeV/a.m.u. deposits the majority of its energy into the electronic system of the material, exciting so-called delta-electrons with energies up to a few tens of keV and creating core holes on attosecond time scales [7–9]. Those delta-electrons scatter further on the electronic and ionic systems of the target. The scattering on the electronic system, referred to as "*inelastic scattering*", includes secondary impact ionization, interaction with the valence and conduction-band electrons, and scattering on the electronic collective modes (plasmons). Pairwise electron-atom interactions, as well as interaction with collective atomic modes, constitute the scattering of electrons on the ionic system of a target (often referred to as "*electron-phonon coupling*"). Both processes take place within ~100 fs timescales, until the electronic transport reduces the density of the excited electrons and the density of excess energy in the electronic ensemble around the SHI trajectory to negligible values [9].

Core holes decay *via* Auger and radiation channels, producing secondary electrons or photons, respectively. Typically, the decay cascade occurs within a few femtoseconds [9,10]. By that time, the majority of holes end up in the valence band of the material, where they are mobile and behave similarly to excited electrons, with their own effective mass [9].

The excited and quickly evolving electronic system interacts with the atomic system of the target in two major ways: apart from scattering and exchanging kinetic energy, it also modifies the atomic potential energy surface. This is known as *nonthermal effects*, the most prominent experimentally observed example of which is the nonthermal melting [11–13]. The atoms may react to those altered interatomic forces by disorder, or – at sufficiently high levels of electronic excitation – by ultrafast acceleration, quickly increasing their kinetic energy on the timescale of a hundred femtoseconds [14].

The atomic system quickly disorders, emitting shock waves and eventually cooling down, typically during about 100 picoseconds [8,9]. During this kinetic stage, the atomic system may recrystallize or remain disordered, depending on the particular material [15,16].

Such a highly nonequilibrium and quickly evolving system poses fundamental interest in many scientific fields, naturally coupling solid-state physics, plasma physics, nonequilibrium kinetics and thermodynamics, radiation transport, and multiple subfields in numerical modelling and simulations [9,17].

## The thermal spike model

To this day, one of the most popular models describing the track radii in various materials remains the inelastic thermal spike (iTS) model [8]. It is based on the two-temperature model (TTM), proposed by Lifshits, Kaganov, and Tanatarov in [18] (relying on the earlier idea of Ginsburg and Shabanskii [19]), which describes the irradiated target with a system of two macroscopic parabolic equations for the coupled temporal and spatial evolution of the thermodynamic temperatures of the electronic and atomic systems:

$$\begin{aligned} c_e(T_e)\frac{\partial T_e}{\partial t} &= \nabla(\kappa_e(T_e,T_i)\nabla T_e) - G(T_e,T_i)(T_e-T_i) + Q_s(\vec{r},t), \\ c_a(T_i)\frac{\partial T_i}{\partial t} &= \nabla(\kappa_i(T_e,T_i)\nabla T_i) + G(T_e,T_i)(T_e-T_i) \end{aligned} \quad (1)$$

where $T_{e,i}$ are the electronic and ionic (atomic) temperatures, respectively; $c_{e,i}$ are the electronic and atomic heat capacities; $\kappa_{e,i}$ are their thermal conductivities; $G$ is the electron-ion (or, as usually assumed, electron-phonon) coupling parameter; and $Q_s$ is a source term of energy delivery into the electronic system by the incoming radiation. It was quickly noticed that applying this model with estimated coefficients to SHI irradiation does not produce sufficient heating in the atomic system to induce observable material modifications [18].

However, this model found appropriate applications in the laser-irradiation community [20,21].

Later, the SHI community revisited the model, noticing that the coefficients in the TTM are not constants but depend on the temperatures, making the system of equations nonlinear [22]. This improvement still did not provide a quantitative agreement of the track size with experiments. It motivated the development of the iTS model, in which it was suggested to use the "electron-phonon coupling" as a free parameter to reproduce the track size (defined in the model as the radius at which the atomic temperature reaches the melting or boiling point) [23]. Alternatively, the source term (the initial conditions for the electronic temperature evolution) may be used as an adjustable parameter to reproduce experimentally observable track radii [24]. This simple empirical approach allows easy estimation of expected track sizes for experiments.

At the same time, it was shown that accounting for temperature dependence in electron-ion coupling reduces the atomic heating rate to values insufficient to induce phase transitions [25]. If it was not possible to describe track formation without the use of a free parameter, the question arises: what is the physics behind the energy exchange between the electronic and atomic systems in an SHI track? Which effects were hidden behind the fitting procedure?

## Rigorous derivation of TTM

Let us revisit the derivation of the TTM from first principles, to make the assumptions and limitations behind it explicit. We will use the generalized temperature approach for that, as it provides a straightforward derivation and allows us to examine each step along the way [26]. Starting with the Liouville equation for the target material under irradiation:

$$\int_{\Omega} \frac{df}{dt} \varphi d\Omega + \int_{\Omega} \sum_{\gamma=1}^{\nu} \sum_{k=1}^{N_\gamma} \left\{ \vec{v}_{\gamma k} \frac{\partial f}{\partial \vec{r}_{\gamma k}} \varphi - \frac{\partial H_{tot}}{\partial \vec{r}_{\gamma k}} \frac{\partial f}{\partial \vec{p}_{\gamma k}} \varphi \right\} d\Omega = 0, \qquad (2)$$

it defines the transport of an arbitrary time-independent quantity, $\varphi$ (which we will use to derive energy transport, following Refs.[27,28]); $f$ is the distribution function; $H_{tot}$ is the Hamiltonian defining the system; $\vec{v}_{\gamma k}$ are the velocities of different sorts of particles ($\vec{v}_{e,a}$ for electrons and atoms/ions in our case); the index *k* indicates summation over various sorts of particles $\gamma$ in the ensemble of $N_\gamma$ particles; $\nu$ stands for the number of different sorts (again, electrons and ions), integrated over the phase space $\Omega$ spanning the coordinate and momentum spaces, $\vec{r}_{\gamma k}$ and $\vec{p}_{\gamma k}$. For electrons and ions, it produces the equation [26]:

$$\int_\Omega \frac{df}{dt}\varphi d\Omega + \int_\Omega \left\{ \left[ \vec{v}_e \frac{\partial f}{\partial \vec{r}_e}\varphi - \frac{\partial H_{tot}}{\partial \vec{r}_e}\frac{\partial f}{\partial \vec{p}_e}\varphi \right] + \left[ \vec{v}_i \frac{\partial f}{\partial \vec{r}_i}\varphi - \frac{\partial H_{tot}}{\partial \vec{r}_i}\frac{\partial f}{\partial \vec{p}_i}\varphi \right] \right\} d\Omega = 0. \tag{3}$$

Using the kinetic and potential energies of electrons and ions ($\varphi = K_e, K_i, U_e, U_i, V_{ei}$), the energy transport equations can be derived in a general case [26]. To arrive at the TTM, the following approximations are required:

(i) It must be assumed that both the electronic and atomic systems are in their local thermodynamic equilibria. That means the electronic distribution function is the equilibrium distribution, encompassing both valence band electrons (holes) and excited conduction band electrons, into the unique distribution function with the same electronic temperature and chemical potential.

Let us clarify the terminology here: any deviation from the thermal equilibrium will be called a *nonequilibrium distribution*, and the system may be referred to as *nonthermalized*. To avoid possible confusion, the term *nonthermal* will be reserved exclusively for the effects of changes in the interatomic potential and forces (nonthermal melting) mentioned above.

The assumption of local thermal equilibrium in the electronic and, separately, atomic (ionic) system means that their global distribution function is a Gibbs-like distribution with two distinct temperatures, $T_e$ and $T_i$ [26]:

$$f = A\, e^{-(K_e+U_e+V_{ei}/2)/T_e} e^{-(K_i+U_i+V_{ei}/2)/T_i}, \tag{4}$$

In this case, Eqs.(3-4) reduce to the following system for electrons and ions [26,29]:

$$\begin{cases} \frac{\partial\langle E_e\rangle}{\partial T_e}\frac{\partial T_e}{\partial t} + \frac{\partial\langle E_e\rangle}{\partial T_i}\frac{\partial T_i}{\partial t} = -\nabla \vec{q}_e + \langle \vec{F}_{ei}\frac{\vec{v}_e+\vec{v}_i}{2}\rangle \\ \frac{\partial\langle E_i\rangle}{\partial T_e}\frac{\partial T_e}{\partial t} + \frac{\partial\langle E_i\rangle}{\partial T_i}\frac{\partial T_i}{\partial t} = -\nabla \vec{q}_i - \langle \vec{F}_{ei}\frac{\vec{v}_e+\vec{v}_i}{2}\rangle \end{cases}, \tag{5}$$

with the total energies in each system denoted as $E_{e,i} = K_{e,i} + U_{e,i} + V_{ei}/2$; the forces acting between electrons and ions are $\vec{F}_{ei}$; the total heat flows $\vec{q}_{e,i} = \vec{q}^{K}_{e,i} + \vec{q}^{U}_{e,i} + 1/2\vec{q}^{U}_{ei}$; the force-velocity correlator is the general form of the electron-ion coupling strength: $G(T_e, T_i) = \langle \vec{F}_{ei} \frac{\vec{v}_e + \vec{v}_i}{2} \rangle / (T_e - T_i)$; and the angle brackets are used for an ensemble average.

Note that the left-hand side of the equations (5) contains not only the specific heat capacities of electrons and ions, $C_e = \frac{\partial \langle E_e \rangle}{\partial T_e}$ and $C_i = \frac{\partial \langle E_i \rangle}{\partial T_i}$, but also cross terms, responsible for *nonthermal* effects: the dependence of the electronic energy on the ionic temperature, $\frac{\partial \langle E_e \rangle}{\partial T_i}$, and vice versa, $\frac{\partial \langle E_i \rangle}{\partial T_e}$. The term $\frac{\partial \langle E_i \rangle}{\partial T_e}$ describes exactly the effects mentioned above – the modification of the atomic potential energy induced by the changes in the electron distribution (defined by the change in the electronic temperature under the assumption of the local thermal equilibrium). The term $\frac{\partial \langle E_e \rangle}{\partial T_i}$ is for the changes in the electronic structure (band structure) due to increase in the atomic temperature.

To reduce the system of equations (5) to the classical TTM form (1), two more approximations are required.

(ii) It must be assumed that the atomic energy is independent of the electronic temperature, and vice versa (no nonthermal effects): $\frac{\partial \langle E_e \rangle}{\partial T_i} = \frac{\partial \langle E_i \rangle}{\partial T_e} = 0$, which is known as the weakly-coupled limit in the warm dense matter field [29]. Then the left-hand sides reduce to the familiar form:

$$\begin{cases} C_e \dfrac{\partial T_e}{\partial t} = -\nabla \vec{q}_e + G(T_e, T_i)(T_e - T_i) \\ C_i \dfrac{\partial T_i}{\partial t} = -\nabla \vec{q}_i - G(T_e, T_i)(T_e - T_i) \end{cases}, \tag{6}$$

(iii) The energy fluxes require the approximation of the Fourier law, which assumes diffusive transport of heat and infinite speed of signal propagation, instead of the ballistic one (wave-like, finite propagation speed) [30,31]:

$$\vec{q}_{e,i} = -\kappa_{e,i} \nabla T_{e,i}, \tag{7}$$

Only then we arrive at the TTM system of equations (1), which, chosen a particular shape of the added source term and fitting procedure for the coupling parameter, becomes the inelastic thermal spike model.

Systematic research over the past few decades has demonstrated that those three assumptions are not satisfied in an SHI track [9]. The energy transport is not diffusive and cannot be described by the Fourier equation in both electronic and atomic systems [32,33]. The electrons are not in equilibrium; their distribution function does not follow the Fermi-Dirac distribution within the time of electron cooling; in particular, the valence holes and electrons behave very differently at those times [34,35]. The weakly-couped limit does not hold in SHI tracks: levels of electronic excitation are high above the nonthermal damage threshold, implying that nonthermal effects may not be neglected in the track center [14,36].

## Theory beyond TTM

To describe the track formation without fitting parameters, it thus requires going beyond the discussed assumptions. The three abovementioned classes of effects must necessarily be accounted for: non-diffusive transport of energy in the electronic and atomic systems; nonequilibrium distributions in the electronic and atomic ensembles and ensuing nonequilibrium energy exchange between the systems; and nonthermal atomic acceleration, caused by transient changes in the interatomic forces.

Currently, there is no single method that could incorporate all those effects [17]. For instance, *ab-initio* simulations can precisely model nonthermal melting and acceleration effects, but they are not capable of tracing nonequilibrium electron dynamics and sufficiently large systems to describe gradients in the track formation problem and ensuing transport effects (characteristic size of millions of atoms and timescales of hundreds of picoseconds). Classical molecular dynamics simulations may reach the necessary time and spatial scales and describe nonequilibrium kinetics in the atomic system, but do not include information about the electronic transport. Whereas methods tracing nonequilibrium electron kinetics, such as the Boltzmann equation or transport Monte Carlo simulations, do not describe the nonthermal melting effects and atomic response [17].

The only alternative proposed so far is a combined, or multiscale, approach, which integrates all the necessary elements from the relevant methods to cover the entire process of track formation [9,17,37]. Combining different methods in the case of an SHI track may be simplified, considering the time-separability of the processes involved: the majority of the electronic effects are over by the time of their cooling (defined by the electronic transport, ~100 fs), whereas the atomic response takes place after that, reacting to the energy deposition from the electrons, and following cooling and recrystallization (defined by the energy and mass transport in the atomic system, ~100 ps).

Below, we discuss an example of such an approach. The electronic transport may be described with classical transport Monte Carlo codes, with necessary modifications described below [34,37]. The nonthermal effects may also be estimated from the data obtained from the electronic transport, using the energy conservation constraints and the equipartition theorem [9,37]. The atomic system may then be described with classical molecular dynamics, with a suitable potential – independent of the electronic temperature since the electronic system is already cold by the time of atomic reaction [37].

### *Nonequilibrium electron transport*

Transient evolution of the electronic system excited by a swift heavy ion may be well described with a transport Monte Carlo method [38–40]. The electronic system effectively splits into a few interdependent subsystems: fast electrons (delta and secondary, with energies above some 10 eV) exhibit strongly non-equilibrium and ballistic transport, gradually turning into more diffusive transport as they lose energy in cascades; slow electrons, forming the majority of electrons in the track center, which are closer to thermal equilibrium and diffusive behavior; valence-band electrons, which may be described as an ensemble of holes with their own effective mass.

The fast electrons are best approximated as classical particles, whose transport may be described by random sampling of scattering events and free flight between them. The scattering on the electronic system (secondary electron ionization, plasmon creation), scattering on the atomic system without its ionization, which may include individual and collective interaction with atoms, and Bremsstrahlung emission are the main interaction channels for this fraction of electrons [37,38].

In this framework, the probabilities of all events are evaluated from the cross sections of scattering. For fast electrons, one may use first-order perturbation theory to evaluate those cross sections, also known as the linear response theory, or the Born approximation [9,38]. Sampling each scattering event of each individual particle allows tracing their nonequilibrium kinetics and natural transition from the initial ballistic to progressively more diffusive transport [41].

Slow electrons can also be modelled with the transport Monte Carlo method, if appropriate cross sections are used accounting for screening effects in the material [38,39]. It is especially the case for modelling SHI impacts, since in this case the time span is naturally limited by ~100 fs, during which electrons perform only a few collisions, and thus their distribution does not accumulate large errors before the simulation is stopped.

### *Valence hole transport*

All core holes decay *via* a cascade of Auger and radiative processes, including such solid-state effects as interatomic Auger (also known as Knotek-Feibelman decay, or interatomic Coulombic decay, and a class of related processes involving electrons from different atoms in molecules and solids) [42,43]. This quickly brings all the holes into the valence band, where they are free to move and scatter, similar to the excited electrons in the conduction band. Depending on their energy and the material band gap, holes may perform secondary impact ionizations and elastic scattering on atomic displacements (valence hole-phonon coupling). These processes may also be simulated with the transport Monte Carlo method, provided effective mass and appropriate scattering cross sections are used [34,39].

### *Nonthermal effects*

Transient changes in the interatomic potential due to electronic excitation may accelerate atoms at extremely fast timescales [13]. This may result in ultrafast atomic disordering, known as nonthermal melting, and form a distinct channel of energy transfer between excited electrons and atoms [14]. This effect is associated with rearrangement of the electronic structure (band structure), in particular leading to the closure of the band gap (band gap collapse) [14,35]. Electrons, occupying the energy levels above the band gap, are thus shifted towards the top of the valence band (and, correspondingly, holes occupying the top of the valence band are shifted towards the conduction band). The change in the potential energy in the electron-hole pairs is converted into the kinetic energy of atoms [14].

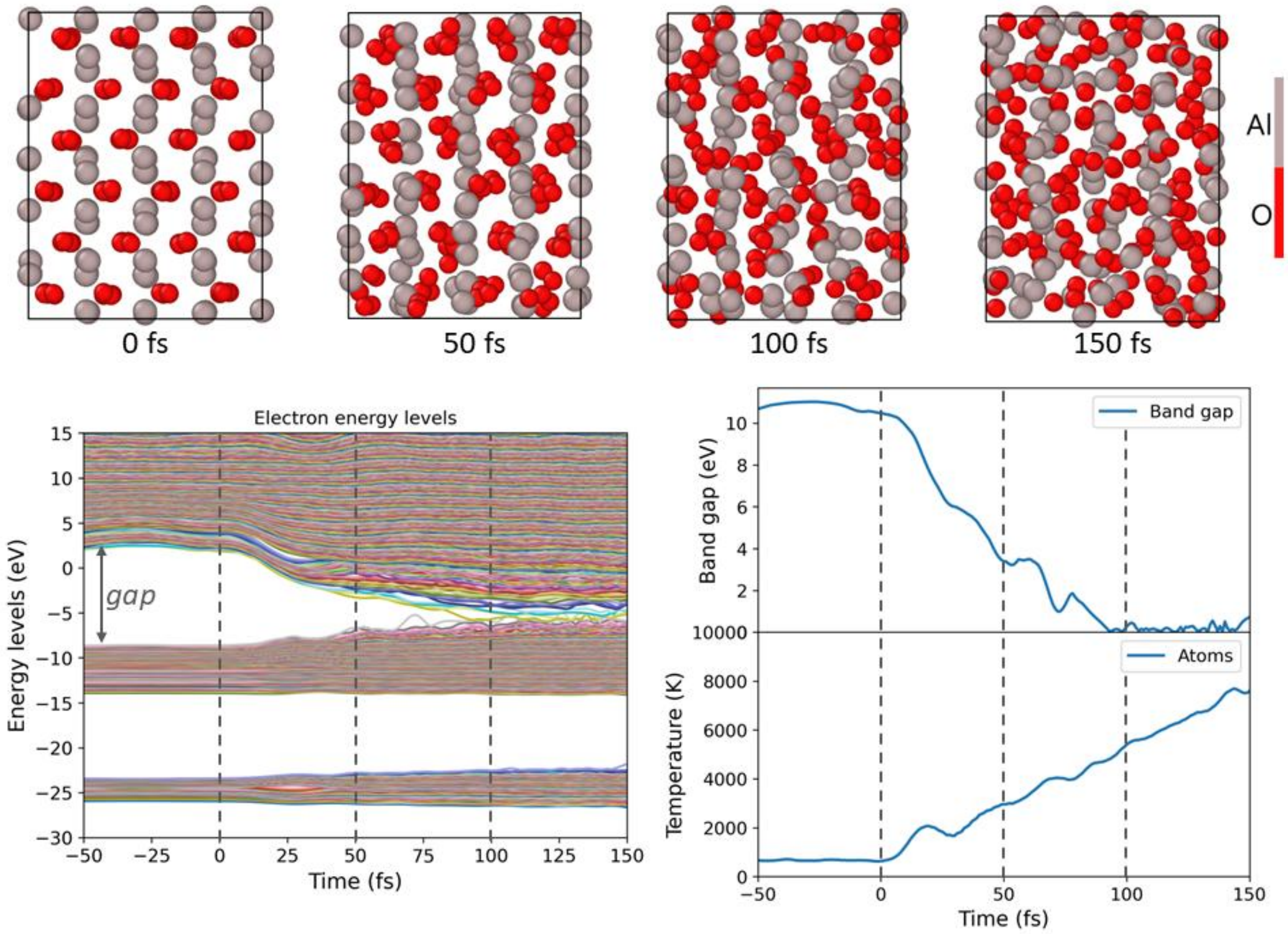

*Figure 2. Example of the response of $Al_2O_3$ to deposition of 6 eV/atom energy into its electronic system at time 0 fs. Top panel: atomic snapshots at different times created with help of OVITO [44]. Bottom left panel: evolution of the electronic energy levels (molecular orbitals, band structure). Bottom right panel: corresponding band gap, and the atomic temperature. Vertical dashed lines mark the time instants corresponding to the atomic snapshots shown. Calculated with XTANT-3 [45].*

These effects may be modelled precisely with *ab-initio* methods, such as TDDFT, DFT-MD, or tight-binding-MD [46–50]. However, such methods are limited to small systems, currently of about 1000 atoms due to computational resources consumption.

An example of such nonthermal effects in $Al_2O_3$ in response to excitation of its electronic system with the dose of 6 eV/atom is shown in Figure 2. Simulation with XTANT-3 demonstrates that exciting the electronic system with the dose above ~4 eV/atom triggers ultrafast atomic acceleration, associated with the rearrangement of the electronic structure of the material [14]. The electronic energy levels (molecular orbitals, band structure) start shifting, bands broadening, with extensive lowering of the conduction band towards the top of the valence band. This rearrangement, thus, manifests as the bandgap collapse, turning the material transiently electronically conducting. The electron-hole pairs, populating the bottom of the conduction band and the top of the valence band, meet, and the associated energy is released into the atomic system, accelerating atoms – notice a sudden jump in the atomic temperature in Figure 2. In response, the atoms disorder on the timescales of ~100 fs, within the characteristic time of the electronic excitation in an SHI track [14].

An approximate method has recently been proposed and validated, suggesting the use of energy conservation and the equipartition theorem to approximately take into account the atomic acceleration caused by nonthermal effects in an SHI track [37]. Taking into account that the band gap collapse is associated with the nonthermal atomic acceleration, the total energy transfer from the electronic to the atomic system due to these effects may be approximately estimated by the number of produced electron-hole pairs and the band gap energy: $E_{nt} \approx E_{gap} n_e$ [37].

This amount of energy, whose distribution in space can be evaluated from the transport Monte Carlo, may then be delivered to the atoms as additional *kinetic* energy, instead of modelling exact atomic acceleration by the change in the *potential* energy – as long as the amount of the transferred energy is identical [37]. This approximation relies on the equipartition theorem, which states that the kinetic and potential energy will be automatically shared equally within the time of atomic thermalization (typically, the timescale of a few atomic oscillations) [26,51]. Essentially, it means that the system loses its memory about how exactly the energy was delivered to it, once thermalized.

This justifies the approach in which the energy transferred from electrons to atoms is used as initial conditions for molecular dynamics simulation. Such an approach produces nearly identical results to those where the energy is delivered gradually, and to the case of *ab-initio* simulation of the nonthermal acceleration [37], at the price of being unable to trace the initial phase of the atomic heating – only the atomic dynamics starting from ~100 fs could be described, including the final track size and structure.

### ***Numerical approach***

Combining the effects described above, the following methodology is employed in the simulation of SHI track formation without fitting parameters, based on the Monte Carlo simulation tool TREKIS-3 for electron kinetics, and the molecular dynamics method for atomic dynamics (e.g., LAMMPS [52]).

TREKIS-3 is a transport Monte Carlo code that simulates SHI impact on matter [34,53]. It uses an event-by-event modelling scheme, in which propagation of each particle is traced. It starts with the incoming ion, sampling individual impact ionization events, exciting delta-electrons, and creating core holes. Then, the holes decay via Auger or radiative channels, emitting secondary electrons or photons, respectively. All electrons undergo their own transport, scattering elastically (without ionization) or inelastically (impact ionization, or plasmon creation). Kinetics of the valence band holes is modelled in the same manner, with their own effective mass, calculated from the density of states of the material within an effective one-band approximation [54]. Photons are transported until photoabsorption takes place, which emits secondary electrons.

The ionization potentials of all shells of all chemical elements of the target, as well as their characteristic Auger and radiative decay times, are extracted from the EPICS2025 database [55].

For the cross section of charged particle scattering, the Ritchie and Howie method is used, which relies on the linear response theory [56]. In this method, the loss function in the cross sections is reconstructed from the optical data. For the swift-heavy ion impact ionization cross section, the Barkas effective charge of the projectile is used [57]. For electron-atom scattering, the cross section includes dynamical screening of the target atomic system [58].

An example of the simulation of 167-MeV Xe ion impact on $Al_2O_3$ (10 Å thick layer simulated with periodic boundary conditions along the boundaries perpendicular to the ion impact trajectory), performed with TREKIS-3, demonstrates the above-mentioned nonequilibrium effects: there is a pronounced non-diffusive front of propagation of the electronic excitation; see Figure 3. Additionally, there is a forefront formed by the photoelectrons. Behind the delta-electron front, there is a second front formed by plasmon formation and decay, which provides secondary non-diffusive transport. Only the fraction of slow secondary electrons exhibits a plateau near the track center (the ion trajectory).

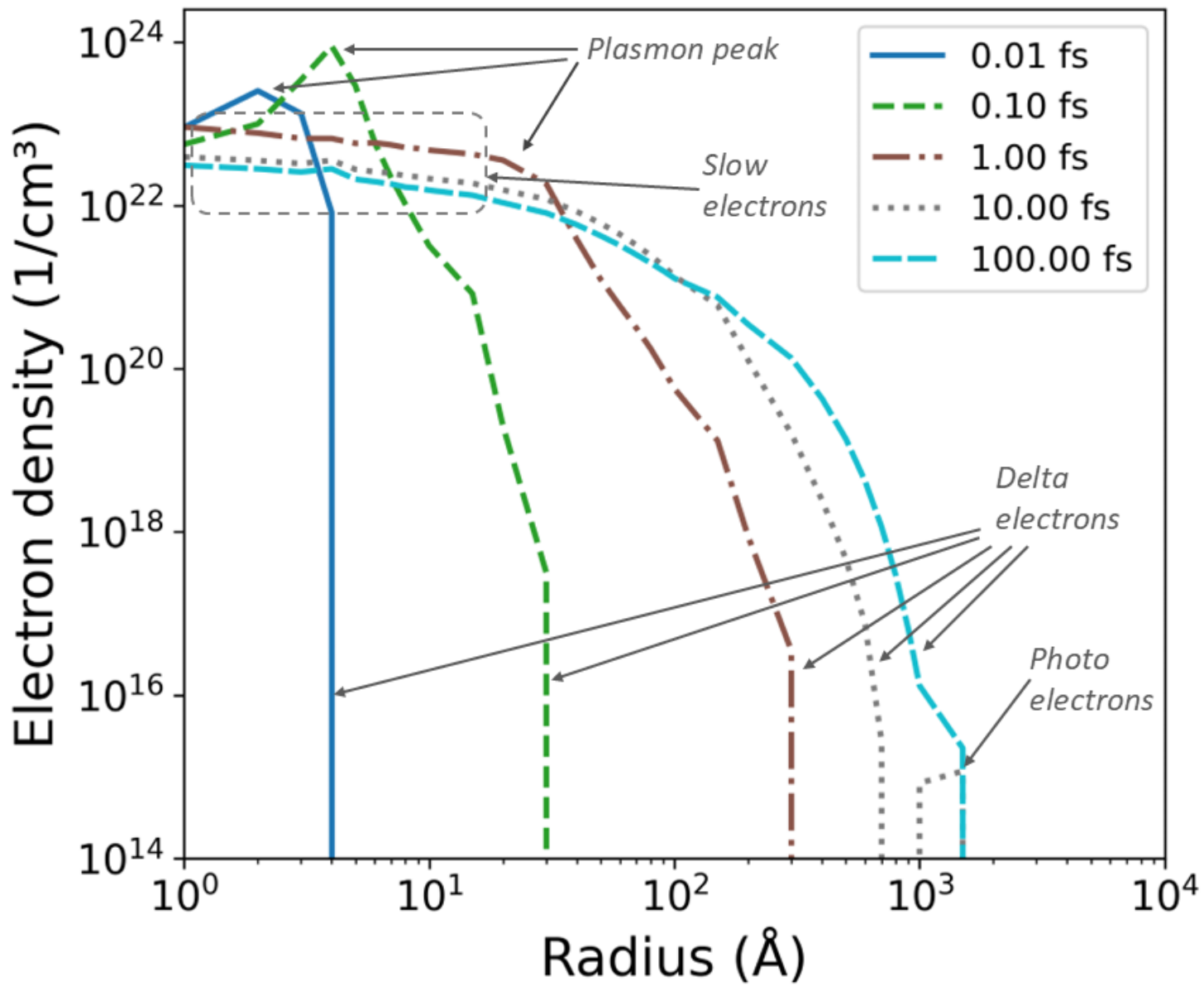

*Figure 3. Radial distribution of electron density in $Al_2O_3$ irradiated by 167 MeV Xe ion.*

At the same time, a significant amount of energy is stored in the valence-band hole system (including electron-hole pair potential energy, associated with the material band gap energy); see the transient energy balance in Figure 4. Both electrons and holes scattering on the atomic lattice deliver kinetic energy to atoms. However, a large portion of the energy resides in the electron-hole pairs; this portion of energy will be delivered to atoms *via* nonthermal acceleration (see previous Section, Figure 2).

Combining these sources of energy, the energy density delivered to the atomic system by the time of the electronic cooling (~100 fs) forms the following radial distribution; see Figure 5. The energy associated with the nonthermal effects, as described above, forms a significant contribution to the atomic heating. We can also see that the energy in the electron-hole pairs is significantly above the nonthermal acceleration threshold (discussed in the last Section), confirming that the nonthermal acceleration is expected to take place in an SHI track, validating the approximation used in the model.

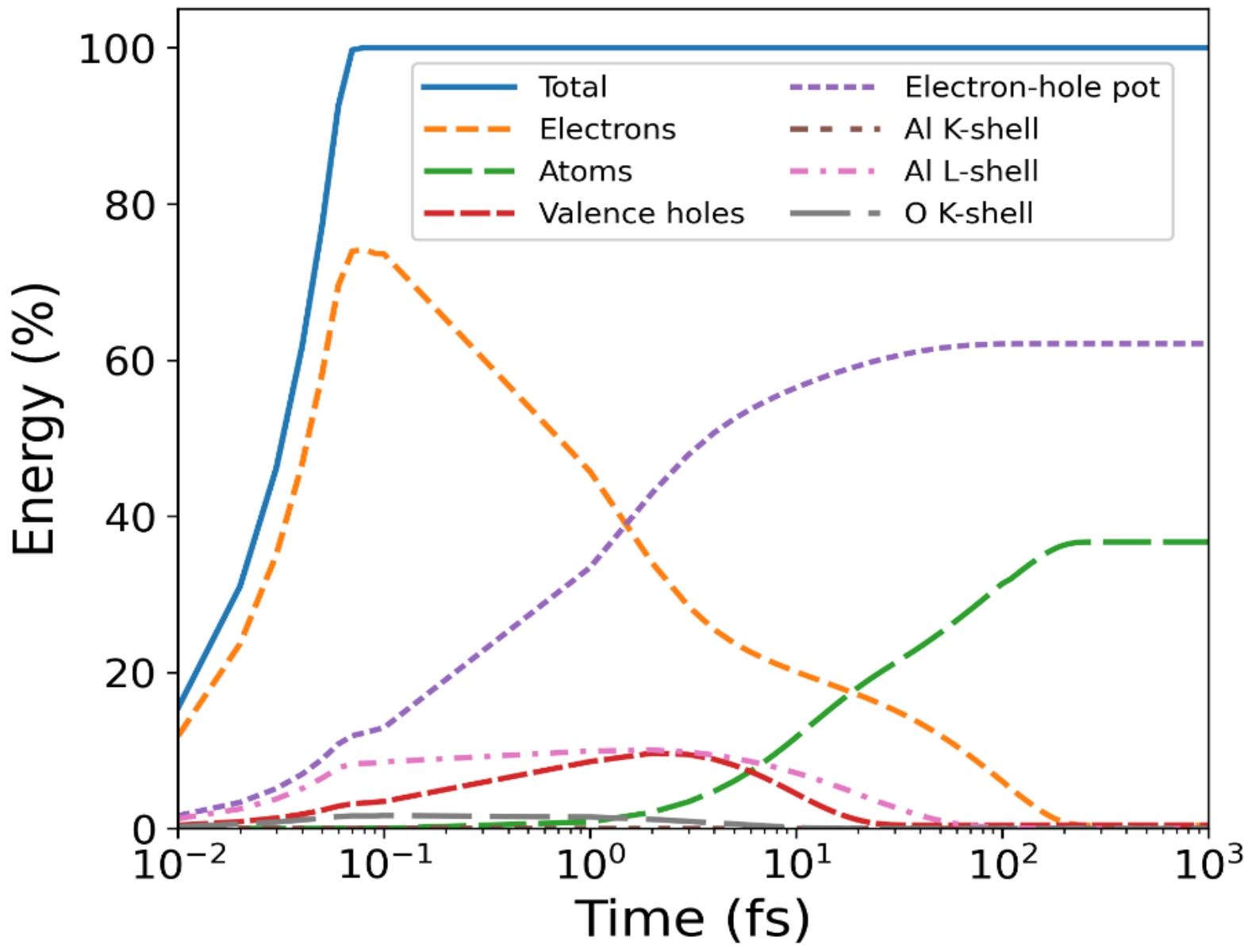

*Figure 4. Fractions of energy in various systems in a 10 Å thick layer of $Al_2O_3$ irradiated by 167 MeV Xe ion. Total energy is defined by the deposition by SHI; the shell-resolved hole energies are shown for each element and the valence band; the electron-hole potential energy is associated with the bandgap energy and the number of excited electrons and holes.*

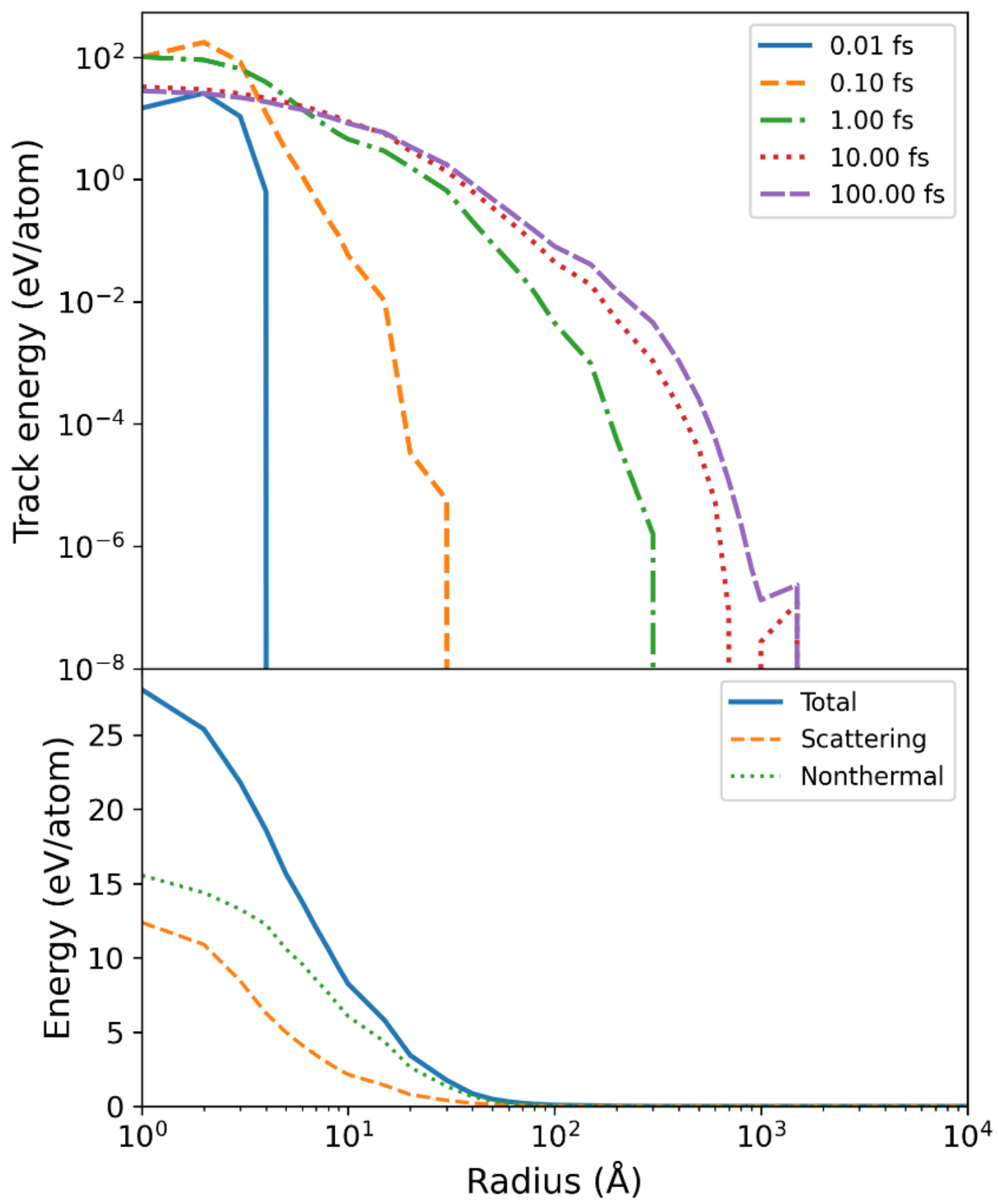

*Figure 5. Radial distribution of the energy density in the atomic system in $Al_2O_3$ irradiated by 167 MeV Xe ion. Top panel: the evolution of the total atomic energy density. Bottom panel: the decomposition of the atomic energy density at 100 fs into*

*the contributions from the scattering of electrons and valence holes, and from the nonthermal effects (associated with the bandgap energy in the electron-hole pairs).*

As was studied in detail in Ref. [37], if the correct amount of energy is delivered to the atoms as the initial kinetic energy, the atomic system quickly relaxes to the identical conditions to those achieved by transient modification of the potential energy. This comparison validated the simplified model of using the TREKIS-3 output of the atomic heating (e.g. Figure 5) as initial conditions for the atomic velocities in further molecular dynamics simulation of the atomic response.

It was proven that such a combination of the MC and MD simulations can describe the track size in good agreement with experimental data without fitting parameters [16,59,60].

However, note that to achieve this, the MC code must account for the three key effects: (i) the electronic scattering with appropriate cross sections tacking into account collective response of the atomic and electronic systems of a target to excitation; (ii) valence hole transport, including scattering on the atomic system; (iii) nonthermal effects, which may be approximated from the MC distribution of electron-hole pairs and the bandgap of the material.

## Limitations of the multiscale approach

The approach described above, despite overcoming the approximations of the two-temperature model, has its own limitations. First of all, a sequential combination of the Monte Carlo method describing the electron kinetics and the molecular dynamics tracing the atomic response (with initial conditions defined by the energy deposition from MC) completely omits the stage of the concurrent electronic and atomic kinetics [37]. This approach is valid for the description of the final track size and structure, but fails at describing the femtosecond atomic dynamics. If experimental methods, e.g., similar to pump-probe techniques, could access the transient states during such ultra-short timescales, a different method would be required, concurrently tracing the evolution of the electronic and atomic dynamics. Such an option was studied with the TREKIS-4 code, demonstrating that the difference between the concurrent electron-atom dynamics and the instantaneous energy deposition into the atomic system washes out during a few hundred femtoseconds [37].

Another bottleneck of the combined MC-MD model is the fact that the band-gap collapse, associated with the nonthermal effects, was only shown to take place in covalent materials. In ionic crystals, only a partial band gap shrinkage occurs at high deposited doses [50]. Therefore, the approximation connecting the electron-hole pair density with the deposited energy *via* the bandgap fails in this case, and a more detailed *ab-initio* estimation of the atomic acceleration is required.

In metals, the nonthermal effects look very different from both above-mentioned cases, where electronic excitation may lead to the opposite effect: phonon hardening, which transiently stiffens the interatomic potential, increasing the atomic barriers and melting point [61]. Absence of nonthermal acceleration of atoms may be one of the reasons why SHI tracks typically do not form in metals.

We must also note that nonthermal melting and atomic acceleration, even in covalent materials, are threshold processes. That means not the entire track will have sufficient electronic excitation to cause it – both in depth and in radius from the SHI trajectory. Notice that the threshold for ultrafast nonthermal melting in Figure 5 is reached only within the radius of ~2 nm. Thus, one should account for the fact that the energy transfer to the atoms due to nonthermal effects is spatially inhomogeneous, and the approximation of the energy source defined by the bandgap requires further correction, albeit apparently relatively small [62].

It was already mentioned above that the MC simulation treats electrons and valence band holes as classical particles described within the linear response theory. Even though it proved sufficient to achieve agreement with experimental track sizes, this approximation may further be improved by replacing the treatment of slow particles with quantum mechanical methods, e.g., similar to the XTANT-3 simulation of laser irradiation [45].

Furthermore, the cross sections of scattering, as well as valence-hole effective mass, are calculated for unirradiated material. Both quantities could change upon excitation in an SHI track, which would alter the particles' behavior. This effect is expected to be of minor importance, since most of the propagation of the excited electrons and holes takes place at the front, experiencing unirradiated material, and the simulation times are limited to ~100 fs, so the accumulated errors due to material changes would not be large. Nonetheless, this point may be improved in future work.

Last, MC simulations do not account for transient electromagnetic fields induced by the charge separation between travelling electrons and holes, which potentially may accelerate the valence holes' transverse transport, or slow down electrons. This effect is also expected to be small and not to influence the radial distribution of the energy deposition into the lattice significantly, due to short timescales and quick cooling of electrons and holes.

Despite these limitations and approximations, the model demonstrated capabilities of describing track sizes and structure in reasonable agreement with experiment, considering that it does not use any adjustable parameters and has full predictive power [16,59,60]. The points discussed suggest potential directions for future research and improvements in the SHI track modelling.

## Conclusions

This brief review systematically demonstrated where the two-temperature model approximations break down in the description of swift-heavy ion track formation, and how one can go beyond these approximations to describe processes within an ion track. We identify three key effects that must be accounted for: nonequilibrium and non-diffusive electron transport and energy exchange with the atomic lattice; the transport of valence holes and their coupling to the atomic system; and nonthermal acceleration of atoms due to transient modification of the interatomic potential induced by extreme electronic excitation. By disentangling the various processes that play a role within the track, it became possible to develop new models specifically tailored to describe each of these effects.

We outlined a simple model without any empirical fitting to SHI tracks, combining the transport Monte Carlo simulation providing initial conditions for molecular dynamics modelling of the atomic response to an SHI passage. We also discussed drawbacks and limitations of this combined approach, which could be addressed in future work.

## Acknowledgements

Computational resources were provided by the e-INFRA CZ project (ID:90254), supported by the Ministry of Education, Youth and Sports of the Czech Republic. NM thanks the financial support from the Czech Ministry of Education, Youth, and Sports (grant nr. LM2023068), and from the European Commission Horizon MSCA-SE Project MAMBA [HORIZON-MSCA-SE-2022 GAN 101131245].

## AI use declaration

Schematic illustration in Figure 6 was created by Microsoft Copilot AI with iterative prompting based on the review [9]. The authors carefully evaluated the figure for correctness of scientific content and clarity and modified it wherever necessary.

## Data and code availability statement

The codes TREKIS-3, TREKIS-4, and XTANT-3 are publicly available from [45,53,63].

## References

[1] A. Ganguly, S. Sankar Nath, M. Choudhury, Effect of Cu doping and swift heavy ion irradiation on PbS quantum dots and their applications in solar cells, IET Optoelectronics 13 (2019) 113–117. https://doi.org/10.1049/iet-opt.2018.5008.

[2] K.K. Jana, V.K. Tiwari, D.K. Avasthi, T.K. Paine, P. Maiti, New Generation Fuel Cell Membrane Using Swift Heavy Ions, ChemistrySelect 2 (2017) 6413–6437. https://doi.org/10.1002/slct.201700690.

[3] C. Cai, X. Fan, J. Liu, D. Li, T. Liu, L. Ke, P. Zhao, Z. He, Heavy-Ion Induced Single Event Upsets in Advanced 65 nm Radiation Hardened FPGAs, Electronics 8 (2019) 323. https://doi.org/10.3390/electronics8030323.

[4] U. Amaldi, Evolution of hadron therapy from 1935 to 2005: a personal view, Health Technol. 14 (2024) 823–832. https://doi.org/10.1007/s12553-024-00894-z.

[5] S.S. Dahiwale, N.S. Shinde, D. Kanjilal, V.N. Bhoraskar, S.D. Dhole, Swift heavy ion induced single event upsets in high density UV-EPROM's, Nuclear Instruments and Methods in Physics Research Section B: Beam Interactions with Materials and Atoms 266 (2008) 1729–1733. https://doi.org/10.1016/j.nimb.2007.12.047.

[6] V.S. Kokhan, M.I. Dobynde, The Effects of Galactic Cosmic Rays on the Central Nervous System: From Negative to Unexpectedly Positive Effects That Astronauts May Encounter, Biology 12 (2023) 400. https://doi.org/10.3390/biology12030400.

[7] F.F. Komarov, Nano- and microstructuring of solids by swift heavy ions, Physics Uspekhi 187 (2017) 465–504. https://doi.org/10.3367/UFNr.2016.10.038012.

[8] M. Lang, F. Djurabekova, N. Medvedev, M. Toulemonde, C. Trautmann, 1.15 - Fundamental Phenomena and Applications of Swift Heavy Ion Irradiations, in: R.J.M. Konings, R.E.

Stoller (Eds.), Comprehensive Nuclear Materials (Second Edition), Elsevier, Oxford, 2020: pp. 485–516. https://doi.org/10.1016/B978-0-12-803581-8.11644-3.
[9] N. Medvedev, A.E. Volkov, R. Rymzhanov, F. Akhmetov, S. Gorbunov, R. Voronkov, P. Babaev, Frontiers, challenges, and solutions in modeling of swift heavy ion effects in materials, J. Appl. Phys. 133 (2023) 100701. https://doi.org/10.1063/5.0128774.
[10] O. Keski-Rahkonen, M.O. Krause, Total and partial atomic-level widths, Atomic Data and Nuclear Data Tables 14 (1974) 139–146. https://doi.org/10.1016/S0092-640X(74)80020-3.
[11] C.W. Siders, A. Cavalleri, K. Sokolowski-Tinten, Cs. Tóth, T. Guo, M. Kammler, M.H. von Hoegen, K.R. Wilson, D. von der Linde, C.P.J. Barty, Detection of Nonthermal Melting by Ultrafast X-ray Diffraction, Science 286 (1999) 1340–1342. https://doi.org/10.1126/science.286.5443.1340.
[12] A. Rousse, C. Rischel, S. Fourmaux, I. Uschmann, S. Sebban, G. Grillon, P. Balcou, E. Förster, J.P. Geindre, P. Audebert, J.C. Gauthier, D. Hulin, Non-thermal melting in semiconductors measured at femtosecond resolution, Nature 410 (2001) 65–68. https://doi.org/10.1038/35065045.
[13] N.J. Hartley, J. Grenzer, L. Huang, Y. Inubushi, N. Kamimura, K. Katagiri, R. Kodama, A. Kon, W. Lu, M. Makita, T. Matsuoka, S. Nakajima, N. Ozaki, T. Pikuz, A.V. Rode, D. Sagae, A.K. Schuster, K. Tono, K. Voigt, J. Vorberger, T. Yabuuchi, E.E. McBride, D. Kraus, Using Diffuse Scattering to Observe X-Ray-Driven Nonthermal Melting, Phys. Rev. Lett. 126 (2021) 015703. https://doi.org/10.1103/PhysRevLett.126.015703.
[14] N. Medvedev, A.E. Volkov, Nonthermal acceleration of atoms as a mechanism of fast lattice heating in ion tracks, J. Appl. Phys. 131 (2022) 225903. https://doi.org/10.1063/5.0095724.
[15] R.A. Rymzhanov, N. Medvedev, J.H. O'Connell, A.J. van Vuuren, V.A. Skuratov, A.E. Volkov, Recrystallization as the governing mechanism of ion track formation, Scientific Reports 9 (2019) 3837. https://doi.org/10.1038/s41598-019-40239-9.
[16] R.A. Rymzhanov, N. Medvedev, J.H. O'Connell, V.A. Skuratov, A. Janse van Vuuren, S.A. Gorbunov, A.E. Volkov, Insights into different stages of formation of swift heavy ion tracks, Nuclear Instruments and Methods in Physics Research Section B: Beam Interactions with Materials and Atoms 473 (2020) 27–42. https://doi.org/10.1016/j.nimb.2020.04.005.
[17] T. Apostolova, E. Artacho, F. Cleri, M. Cotelo, M.L. Crespillo, F. Da Pieve, V. Dimitriou, F. Djurabekova, D.M. Duffy, G. García, M. García-Lechuga, B. Gu, T. Jarrin, E. Kaselouris, J. Kohanoff, G. Koundourakis, N. Koval, V. Lipp, L. Martin-Samos, N. Medvedev, A. Molina-Sánchez, D. Muñoz-Santiburcio, S.T. Murphy, K. Nordlund, E. Oliva, J. Olivares, N.A. Papadogiannis, A. Redondo-Cubero, A. Rivera De Mena, A.E. Sand, D. Sangalli, J. Siegel, A.V. Solov'yov, I.A. Solov'yov, J. Teunissen, E. Vázquez, A.V. Verkhovtsev, S. Viñals, M.D. Ynsa, Tools for investigating electronic excitation: experiment and multi-scale modelling, Universidad Politécnica de Madrid. Instituto de Fusión Nuclear Guillermo Velarde, Madrid, 2021. https://doi.org/10.20868/UPM.book.69109.
[18] I.M. Lifshits, M.I. Kaganov, L.V. Tanatarov, On the theory of radiation-induced changes in metals, Journal of Nuclear Energy. Part A. Reactor Science 12 (1960) 69–78. https://doi.org/10.1016/0368-3265(60)90010-4.
[19] V.L. Ginzburg, V.P. Shabanskii, On the theory of kinetic phenomena in helium II, (1955) 445.

[20] S.I. Anisimov, B.L. Kapeliovich, T.L. Perelman, Electron emission from metal surfaces exposed to ultrashort laser pulses, Zhurnal Eksperimentalnoi i Teoreticheskoi Fiziki 66 (1974) 776–781.
[21] B. Rethfeld, D.S. Ivanov, M.E. Garcia, S.I. Anisimov, Modelling ultrafast laser ablation, J. Phys. D: Appl. Phys. 50 (2017) 193001. https://doi.org/10.1088/1361-6463/50/19/193001.
[22] I.A. Baranov, Y.V. Martynenko, S.O. Tsepelevich, Y.N. Yavlinskiĭ, Inelastic sputtering of solids by ions, Sov. Phys. Usp. 31 (1988) 1015–1034. https://doi.org/10.1070/PU1988v031n11ABEH005646.
[23] M. Toulemonde, C. Dufour, E. Paumier, Transient thermal process after a high-energy heavy-ion irradiation of amorphous metals and semiconductors, Phys. Rev. B 46 (1992) 14362–14369. https://doi.org/10.1103/PhysRevB.46.14362.
[24] G. Szenes, Comparison of two thermal spike models for ion–solid interaction, Nuclear Instruments and Methods in Physics Research Section B: Beam Interactions with Materials and Atoms 269 (2011) 174–179. https://doi.org/10.1016/j.nimb.2010.11.009.
[25] A.E. Volkov, V.A. Borodin, The rate of the electron-ion heat transfer in swift heavy particle tracks in metals, Nuclear Instruments and Methods in Physics Research Section B: Beam Interactions with Materials and Atoms 107 (1996) 172–174. https://doi.org/10.1016/0168-583X(95)01031-9.
[26] N. Medvedev, A.E. Volkov, Multitemperature atomic ensemble: Nonequilibrium evolution after ultrafast electronic excitation, Phys. Rev. E 110 (2024) 024142. https://doi.org/10.1103/PhysRevE.110.024142.
[27] J.H. Irving, J.G. Kirkwood, The Statistical Mechanical Theory of Transport Processes. IV. The Equations of Hydrodynamics, The Journal of Chemical Physics 18 (1950) 817–829. https://doi.org/10.1063/1.1747782.
[28] R.J. Bearman, J.G. Kirkwood, Statistical Mechanics of Transport Processes. XI. Equations of Transport in Multicomponent Systems, The Journal of Chemical Physics 28 (1958) 136–145. https://doi.org/10.1063/1.1744056.
[29] J. Vorberger, D.O. Gericke, Th. Bornath, M. Schlanges, Energy relaxation in dense, strongly coupled two-temperature plasmas, Phys. Rev. E 81 (2010) 046404. https://doi.org/10.1103/PhysRevE.81.046404.
[30] D.S. Chandrasekharaiah, Hyperbolic Thermoelasticity: A Review of Recent Literature, Applied Mechanics Reviews 51 (1998) 705–729. https://doi.org/10.1115/1.3098984.
[31] G. Capriz, K. Wilmanski, P.M. Mariano, Exact and approximate Maxwell-Cattaneo-type descriptions of heat conduction: A comparative analysis, International Journal of Heat and Mass Transfer 175 (2021) 121362. https://doi.org/10.1016/j.ijheatmasstransfer.2021.121362.
[32] N.A. Medvedev, A.E. Volkov, N.S. Shcheblanov, B. Rethfeld, Early stage of the electron kinetics in swift heavy ion tracks in dielectrics, Phys. Rev. B 82 (2010) 125425. https://doi.org/10.1103/PhysRevB.82.125425.
[33] V.P. Lipp, A.E. Volkov, M.V. Sorokin, B. Rethfeld, Kinetics of propagation of the lattice excitation in a swift heavy ion track, Nuclear Instruments and Methods in Physics Research Section B: Beam Interactions with Materials and Atoms 269 (2011) 865–868. https://doi.org/10.1016/j.nimb.2010.11.070.
[34] N.A. Medvedev, R.A. Rymzhanov, A.E. Volkov, Time-resolved electron kinetics in swift heavy ion irradiated solids, J. Phys. D: Appl. Phys. 48 (2015) 355303. https://doi.org/10.1088/0022-3727/48/35/355303.

[35] N. Medvedev, A.E. Volkov, B. Ziaja, Electronic and atomic kinetics in solids irradiated with free-electron lasers or swift-heavy ions, Nuclear Instruments and Methods in Physics Research Section B: Beam Interactions with Materials and Atoms 365 (2015) 437–446. https://doi.org/10.1016/j.nimb.2015.08.063.
[36] R.A. Voronkov, N. Medvedev, R.A. Rymzhanov, A.E. Volkov, Ab-initio modeling of Al2O3 lattice instability under extreme excitation of the electronic system, Nuclear Instruments and Methods in Physics Research Section B: Beam Interactions with Materials and Atoms 435 (2018) 87–92. https://doi.org/10.1016/j.nimb.2018.03.038.
[37] N. Medvedev, F. Akhmetov, R.A. Rymzhanov, R. Voronkov, A.E. Volkov, Modeling Time-Resolved Kinetics in Solids Induced by Extreme Electronic Excitation, Advanced Theory and Simulations 5 (2022) 2200091. https://doi.org/10.1002/adts.202200091.
[38] T.M. Jenkins, W.R. Nelson, A. Rindi, Monte Carlo Transport of Electrons and Photons, Springer US, Boston, MA, 1988. https://doi.org/10.1007/978-1-4613-1059-4.
[39] C. Jacoboni, L. Reggiani, The Monte Carlo method for the solution of charge transport in semiconductors with applications to covalent materials, Rev. Mod. Phys. 55 (1983) 645–705. https://doi.org/10.1103/RevModPhys.55.645.
[40] M.P.R. Waligórski, R.N. Hamm, R. Katz, The radial distribution of dose around the path of a heavy ion in liquid water, International Journal of Radiation Applications and Instrumentation. Part D. Nuclear Tracks and Radiation Measurements 11 (1986) 309–319. https://doi.org/10.1016/1359-0189(86)90057-9.
[41] O. Osmani, N. Medvedev, M. Schleberger, B. Rethfeld, Energy dissipation in dielectrics after swift heavy-ion impact: A hybrid model, Phys. Rev. B 84 (2011) 214105. https://doi.org/10.1103/PhysRevB.84.214105.
[42] M.L. Knotek, P.J. Feibelman, Stability of ionically bonded surfaces in ionizing environments, Surface Science 90 (1979) 78–90. https://doi.org/10.1016/0039-6028(79)90011-6.
[43] L.S. Cederbaum, J. Zobeley, F. Tarantelli, Giant Intermolecular Decay and Fragmentation of Clusters, Phys. Rev. Lett. 79 (1997) 4778–4781. https://doi.org/10.1103/PhysRevLett.79.4778.
[44] A. Stukowski, Visualization and analysis of atomistic simulation data with OVITO–the Open Visualization Tool, Modelling Simul. Mater. Sci. Eng. 18 (2009) 015012. https://doi.org/10.1088/0965-0393/18/1/015012.
[45] N. Medvedev, XTANT-3 [Computer software], https://zenodo.org/records/19810149, (2026). https://doi.org/10.5281/zenodo.19810149.
[46] H.O. Jeschke, M.E. Garcia, K.H. Bennemann, Microscopic analysis of the laser-induced femtosecond graphitization of diamond, Phys. Rev. B 60 (1999) R3701–R3704. https://doi.org/10.1103/PhysRevB.60.R3701.
[47] C. Lian, S.B. Zhang, S. Meng, Ab initio evidence for nonthermal characteristics in ultrafast laser melting, Phys. Rev. B 94 (2016) 184310. https://doi.org/10.1103/PhysRevB.94.184310.
[48] M. Ono, H. Gonome, A. Komiya, First-principles investigation of the relationship between laser irradiation intensity and non-thermal melting of α-quartz, J. Appl. Phys. 138 (2025) 105103. https://doi.org/10.1063/5.0279804.
[49] T. Zier, E.S. Zijlstra, A. Kalitsov, I. Theodonis, M.E. Garcia, Signatures of nonthermal melting, Struct Dyn 2 (2015) 054101. https://doi.org/10.1063/1.4928686.

[50] R.A. Voronkov, N. Medvedev, A.E. Volkov, Dependence of nonthermal metallization kinetics on bond ionicity of compounds, Sci Rep 10 (2020) 13070. https://doi.org/10.1038/s41598-020-70005-1.
[51] J.G. Powles, G. Rickayzen, D.M. Heyes, Temperatures: old, new and middle aged, Molecular Physics 103 (2005) 1361–1373. https://doi.org/10.1080/00268970500054664.
[52] A.P. Thompson, H.M. Aktulga, R. Berger, D.S. Bolintineanu, W.M. Brown, P.S. Crozier, P.J. In ’T Veld, A. Kohlmeyer, S.G. Moore, T.D. Nguyen, R. Shan, M.J. Stevens, J. Tranchida, C. Trott, S.J. Plimpton, LAMMPS - a flexible simulation tool for particle-based materials modeling at the atomic, meso, and continuum scales, Computer Physics Communications 271 (2022) 108171. https://doi.org/10.1016/j.cpc.2021.108171.
[53] N. Medvedev, R. Rymzhanov, A. Volkov, TREKIS-3, (2025). https://doi.org/10.5281/zenodo.17273753.
[54] R.A. Rymzhanov, N.A. Medvedev, A.E. Volkov, Effects of model approximations for electron, hole, and photon transport in swift heavy ion tracks, Nuclear Instruments and Methods in Physics Research Section B: Beam Interactions with Materials and Atoms 388 (2016) 41–52. https://doi.org/10.1016/j.nimb.2016.11.002.
[55] D. Cullen, Electron Photon Interaction Cross Sections (2025) (EPICS2025), (n.d.). https://nuclear.llnl.gov/EPICS/index.html (accessed July 8, 2026).
[56] R.H. Ritchie, A. Howie, Electron excitation and the optical potential in electron microscopy, The Philosophical Magazine: A Journal of Theoretical Experimental and Applied Physics 36 (1977) 463–481. https://doi.org/10.1080/14786437708244948.
[57] W.H. Barkas, Nuclear research emulsions, New York : Academic Press, 1963. http://archive.org/details/nuclearresearche0001bark (accessed October 18, 2026).
[58] N. Medvedev, D.I. Zainutdinov, A.E. Volkov, Non-ionizing cross section of electron scattering on atoms in matter accounting for dynamical screening effect, J. Appl. Phys. 137 (2025) 015903. https://doi.org/10.1063/5.0242080.
[59] R.A. Rymzhanov, N. Medvedev, A.E. Volkov, Damage threshold and structure of swift heavy ion tracks in Al2O3, J. Phys. D: Appl. Phys. 50 (2017) 475301. https://doi.org/10.1088/1361-6463/aa8ff5.
[60] A.A. Peña, N. Medvedev, Creating mixed-phase states in cadmium sulfide by swift heavy ion irradiation, Journal of Alloys and Compounds 1079 (2026) 189943. https://doi.org/10.1016/j.jallcom.2026.189943.
[61] V. Recoules, J. Clérouin, G. Zérah, P.M. Anglade, S. Mazevet, Effect of Intense Laser Irradiation on the Lattice Stability of Semiconductors and Metals, Phys. Rev. Lett. 96 (2006) 055503. https://doi.org/10.1103/PhysRevLett.96.055503.
[62] R. Voronkov, D. Zainutdinov, N. Medvedev, A.E. Volkov, Effect of radially heterogeneous band gap collapse on formation of swift heavy ion tracks in Al2O3, (2026). https://doi.org/10.48550/arXiv.2607.01016.
[63] N. Medvedev, A. Volkov, R. Rymzhanov, R. Voronkov, F. Akhmetov, TREKIS-4 [Computer software], 10.5281/zenodo.17068795, (2025). https://doi.org/10.5281/zenodo.17068795.